# Proximity Induced Chiral Quantum Light Generation in Strain-Engineered WSe$_2$/NiPS$_3$ Heterostructures

Xiangzhi Li,[1] Andrew C. Jones,[1] Junho Choi,[2] Huan Zhao,[1] Vigneshwaran Chandrasekaran,[1] Michael T. Pettes,[1] Andrei Piryatinski,[3] Nikolai Sinitsyn,[3] Scott A. Crooker,[2] Han Htoon[1]*

[1] Center for Integrated Nanotechnologies, Materials Physics and Applications Division, Los Alamos National Laboratory, Los Alamos, NM 87545

[2] National High Magnetic Field Laboratory, Materials Physics and Applications Division, Los Alamos National Laboratory, Los Alamos, NM 87545

[3] Theoretical Division, Los Alamos National Laboratory, Los Alamos, NM 87545

**Quantum light emitters (QEs) capable of generating single photons of well-defined circular polarization could enable non-reciprocal single photon devices and deterministic spin-photon interfaces critical for realizing complex quantum networks.[1,2] To date, emission of such chiral quantum light has been achieved via the application of intense external magnetic fields[3,4] electrical[5-7]/optical[8-10] injection of spin polarized carriers/excitons, or coupling with complex photonic/meta-structures.[1,2] Here we report free-space generation of highly chiral single photons from QEs created in monolayer WSe$_2$ - NiPS$_3$ heterostructures at zero external magnetic field. These QEs emit in the 760-800 nm range with a degree of circular polarization and single photon purity as high as 0.71 and 80% respectively, independent of pump laser polarization. QEs are deterministically created by pressing a scanning probe microscope tip into a two-dimensional heterostructure comprising a WSe$_2$**

**monolayer and a ~50 nm thick layer of the antiferromagnetic (AFM) insulator NiPS$_3$[11-13]. Temperature dependent magneto-photoluminescence studies indicate that the chiral quantum light emission arises from magnetic proximity interactions[14-17] between localized excitons in the WSe$_2$ monolayer[18-21] and the out-of-plane magnetization of AFM defects in NiPS$_3$, both of which are co-localized by the strain field arising from the nanoscale indentations.**

Defined as the process by which an atomically-thin material acquires properties from adjacent materials via quantum mechanical interactions, proximity effects have recently emerged as a means of inducing desirable magnetic, topological, transport, and optical properties in two dimensional systems.[14,17] Strong enhancement of valley Zeeman splitting[15,22-24] and spin-dependent charge transfer[16,25-27] have been observed through the coupling of semiconducting transition metal dichalcogenide (TMD) monolayers to magnetic thin films (e.g. EuS)[15,22] and Van der Waal (VDW) magnets (e.g. CrI$_3$,[16,27] CrBr$_3$,[25,26] Cr$_2$Ge$_2$Te$_2$,[24] Fe$_3$GeTe$_2$[23] ). To date however, few studies have focused on exploiting proximity effects to manipulate the chirality of Quantum Emitters (QEs) in monolayer TMDs.[23,24,27]

In parallel, the study of transition metal phosphorous trichalcogenide TMPX$_3$ (TM = Mn, Ni, Fe, Co; X = S, Se) VDW antiferromagnetic crystals has recently opened up new frontiers for exploring 2D magnetism with applications in spintronic and quantum information technologies.[28] Recent breakthroughs include the discovery of charge-spin correlation,[29] the spin-induced linear polarization of excitons,[11-13,30] the switching of Néel vector under strain,[31] and the slowdown of spin dynamics near the AFM phase transition temperature.[32] The potential of these materials to support magnons, i.e. the collective excitation of electron spins,[33-36] as a means of coupling QEs represents an additional opportunity for the realization of microwave-magnon-spin quantum interfaces.[37] Still, the study of whether TMPX$_3$ materials are capable of modifying the properties of free and localized excitons in 2D TMD materials via the proximity effect remains in its nascent stage.[38-40] Here we show that WSe$_2$/NiPS$_3$ heterostructures locally strained via nano-indentations[41] can support QEs capable of emitting single photons with strong circular polarization at zero magnetic field and independent of excitation polarization. This observation is surprising because NiPS$_3$ display in-plane AFM order,[33-35] and as a result, it is not expected to yield a net out-of-plane magnetization capable of enhancing the valley Zeeman effect and inducing circularly polarized photoluminescence (PL).

WSe$_2$/NiPS$_3$ heterostructures were fabricated using a dry-transfer system in a glove-box. **Figure 1a - c** displays a representative heterostructure consisting of a ~50 nm thick NiPS$_3$ flake and a monolayer thick WSe$_2$ flake stacked atop of a Si wafer coated with a 300 nm thick polymer layer. Wide-field PL images acquired at cryogenic temperatures ($T$=4 K) under 514 nm continuous wave (CW) laser excitation (Figure 1b) shows that PL from the WSe$_2$ monolayer is strongly quenched when coupled to NiPS$_3$. Under $\sigma^+$ and $\sigma^-$ polarized excitation, PL spectra acquired from regions R1 and R2 showed no degree of circular polarization indicating that e-h pairs preferentially pumped non-resonantly into the K$^+$ valley relax equally to both valleys before recombination[42] (Extended Data Fig. 2).

We created localized QEs on the heterostructure using strain-engineering via nanoindentation.[41] By pressing a blunt atomic force microscope probe (~100 nm radius curvature) into the heterostructure, an array of indentations measuring ~250 nm in diameter, 175 nm in depth, and surrounded by a ~100 nm high circular ridge were created (Figure 1c & d). Bright localized PL emission with spectrally narrow peaks is observed at indentation locations (Figure 1b, e).[41] We observe that the chirality of the localized PL emission from nanoindentations exhibits a variety of behaviors. Some indentations (e.g. spot #9 of Figure 1e) show no detectable circularly polarized emission at zero external magnetic field. Notably, other indentations (spots 1,2,3, and 8) exhibit localized PL emission peaks with $\sigma^+$ or $\sigma^-$ polarization (Figure 1.e and Extended Data Fig. 3) with the degree of circular polarization (DCP= $(I_{\sigma^+} - I_{\sigma^-})/(I_{\sigma^+} + I_{\sigma^-})$) varying widely even among the localized PL emission peaks originating from the same indented spot. Analysis of the spectral image of sharp PL peaks from an individual indentation (Extended Data Fig. 4) reveals that these sharp PL peaks originate from independent localized exciton states separated by fifty to a few hundred nm.

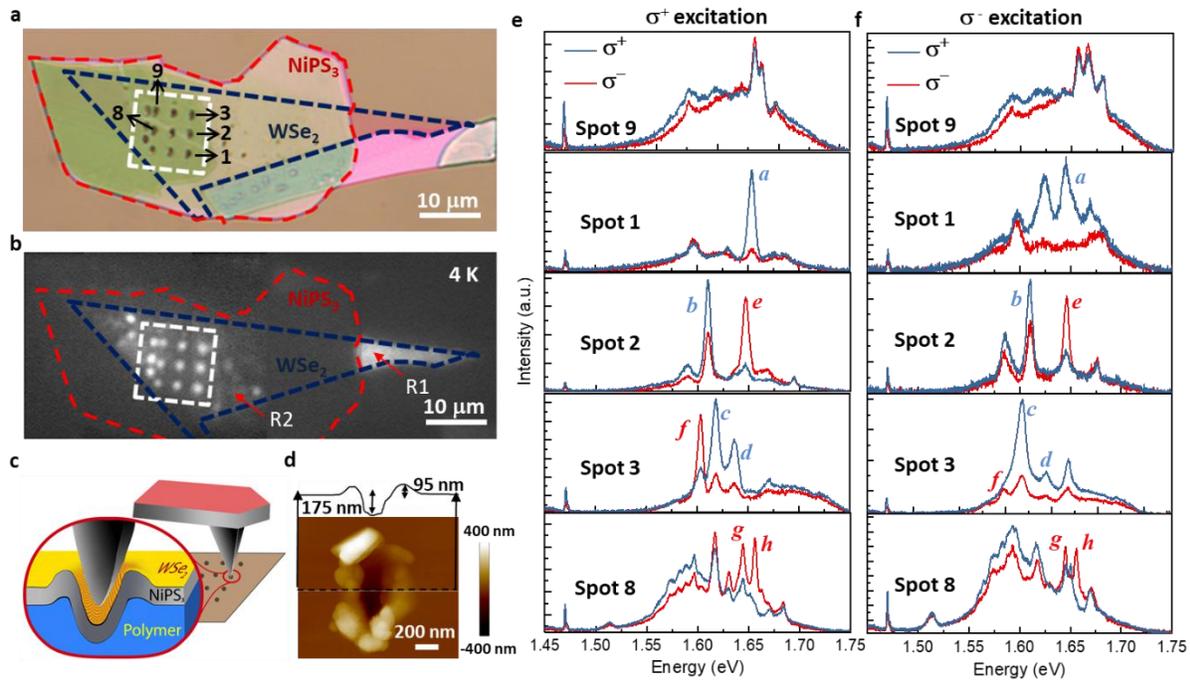

**Figure 1. Strain-engineered WSe$_2$/NiPS$_3$ heterostructures host QEs displaying sharp, localized PL peaks with a strong degree of spontaneous circular polarization. a, b.** Optical **(a)** and photoluminescence **(b)** images of the WSe$_2$/NiPS$_3$ heterostructure. The WSe$_2$ monolayer that does not overlap with NiPS$_3$ emits bright PL(region R1), while PL is quenched when WSe$_2$ is coupled to the NiPS$_3$ (region R2). Strong PL was restored by indentations marked by the white square. **c, d.** Schematic showing sample structure **(c)** with a corresponding atomic force microscopy topography image and cross section of a representative indentation **(d)**. **e, f.** $\sigma^+$ (blue) and $\sigma^-$ (red) resolved, low temperature PL spectra from indentations labeled in **a** acquired

under σ⁺ **(e)** and σ⁻ **(f)** laser excitation. While the peaks marked as *a, b, c & d* in (e) are σ⁺ polarized with DCP of 0.71, 0.48, 0.51 and 0.42 peaks marked *e, f, g & h* are σ⁻ polarized with DCP of -0.56, -0.39, -0.24 and -0.53. All of these localized emission peaks ride on top of a broad PL background that displays little polarization suggesting that DCP of localized emission peaks could even be higher if the broad PL back ground is subtracted. The orientation of the σ$^{+/-}$ polarizations of peaks *a-h* remains the same when the polarization of the laser excitation is switched from σ⁺ **(e)** to σ⁻ **(f)**. However, the DCP of the peaks changes and new PL peaks with varying DCP arise. Sharp emission peak at 1.47 eV has been attributed to anisotropic exciton of $NiPS_3$.[11-13]

To test whether the chiral light emission is independent from the excitation polarization, we reversed the polarization of the laser excitation from σ⁺ to σ⁻. For those localized PL emission peaks exhibiting chiral emission, we observe that while the peak intensity and DCP varied significantly, the overall sign of the chirality was maintained (Figure 1f). In some cases, new PL peaks with varying DCP also arose. Exciting the sample with both circularly and linearly polarized light of different orientation and analyzed the σ⁺ to σ⁻ emission, we observed that the localized PL peaks maintain their overall sign of σ⁺ to σ⁻ polarization for all excitation polarizations (Extended data Fig. 5).

To demonstrate the capability of localized emission sites to act as QEs, we performed Hanbury Brown Twiss (HBT) experiments. **Figure 2a,b** shows circular polarization resolved PL spectra recorded from an indentation excited with either σ⁺ and σ⁻ polarization, respectively. Spectrally filtering the lowest energy peak (E=1.629 eV, DCP= 0.3) showing strong σ⁻ polarization under both excitation for HBT characterization, we next performed time tagged, time correlated single photon counting with two APDs to acquire the intensity-vs-time trace, decay curve, and 2$^{nd}$ order photon correlation function ($g^{(2)}$) of the PL (See methods). The PL intensity vs time trace (Figure 2c) shows blinking-free emission while the decay curve (Figure 2d) can be fit with single exponential with lifetime of 4.35 ± 0.01 ns. Ultimately, the $g^{(2)}$ trace provides clear evidence of quantum light emission with >80% single photon purity (i.e 1- $g^{(2)}(0)/g^{(2)}(T)$) that is comparable to other reported WSe$_2$ QEs.[18-21] We attribute the residual multi-photon emission probability ($g^{(2)}(0)/g^{(2)}(T)$) of 0.2 to the broadband emission tail extending from the group of higher energy PL peaks.

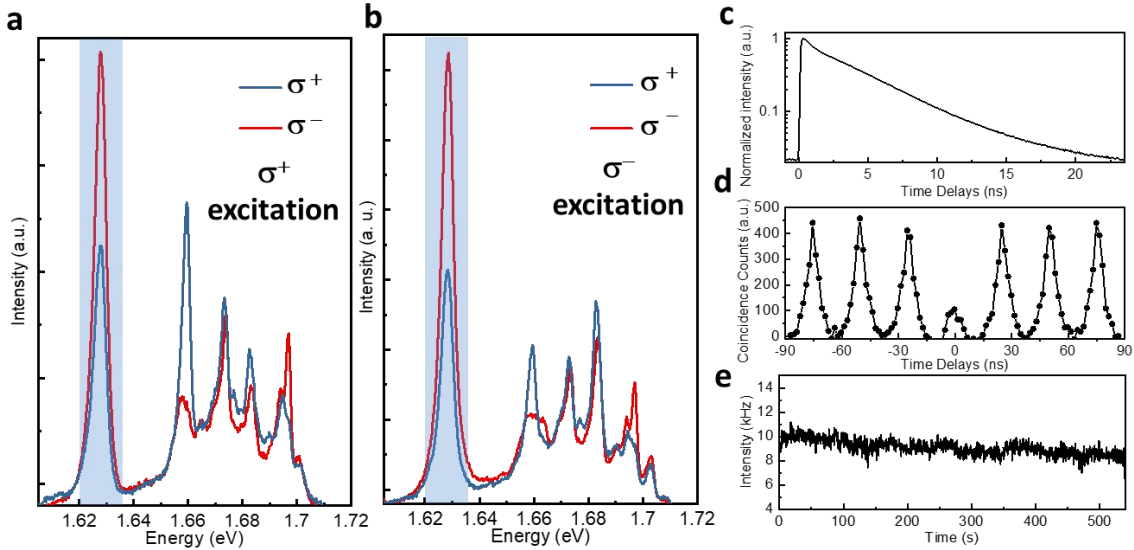

**Figure 2. Hanbury-Brown and Twiss experiments demonstrating quantum light emission from strain-engineered WSe$_2$/NiPS$_3$ heterostructures. a &b.** $\sigma^+$ (blue) and $\sigma^-$ (red) polarized PL spectra of an indented WSe$_2$/NiPS$_3$ heterostructure acquired under $\sigma^+$ **(a)** and $\sigma^-$ **(b)** polarized laser excitation. **c-e.** PL decay curve **(c)**, 2$^{nd}$ order photon correlation trace **(d)**, PL intensity time trace measured from the PL peak marked by blue column in **(a)** and **(b)** acquired under 514 nm, 40 MHz pulsed laser excitation.

Together, these findings provide clear evidence that QEs capable of generating chiral single photons can be created via local strain engineering of WSe$_2$/NiPS$_3$ heterostructures without the need for external magnetic field and independent of laser excitation polarization (Figure 1 & 2, Extended Data Fig. 5). Previously, circularly polarized QEs have been generated via high magnetic fields[18-21,24], via injection of spin polarized carriers,[5-7] and via near-resonant pumping of valley-polarized excitons into donor-bound states[8] and interlayer excitons localized in moiré lattices.[9,10] Spontaneous circular polarization of ligand-field luminescence, defined by out-of-plane magnetization, was recently observed in CrX$_3$ (X= I, Br) monolayers.[43,44] However, due to the localized molecular orbital origin of this luminescence and strong vibronic coupling, emission lines are broad and quantum light emission is unlikely to be attained in such systems.

While recent attempts to induce magnetic proximity effects on TMDs by coupling to EuS,[15,22] CGT,[24] and FGT[23] can provide some enhancement of Zeeman effects, zero-field circularly polarized PL emission has remained elusive to date. Coupling of TMDs to $CrX_3$ gives rise to hysteresis in PL polarization ρ, defined as $\rho(B) = (I_{\sigma-/\sigma-} - I_{\sigma+/\sigma+})/ (I_{\sigma-/\sigma-} + I_{\sigma+/\sigma+})$ following the magnetization of the $CrX_3$ due to spin dependent charge transfer.[26,27] However, localized QEs in such systems exhibit no circular polarization at zero field and only normal Zeeman splitting under external magnetic fields.[26,27] Because $NiPS_3$ exhibits zig-zag AFM order with spins aligned parallel or antiparallel to the in-plane *a* axis compensating each other at the lattice constant (**Figure 3a**),[11-13,33,34] the net magnetization needed to modify optical properties of nearby materials through proximity effects is practically averaged out to zero. Our control experiment on unstrained regions of the $WSe_2/NiPS_3$ heterostructure confirms the absence of net magnetization as the PL emission from $WSe_2$ 2D excitons exhibit no detectable circular polarization (Extended Data Fig. 2).

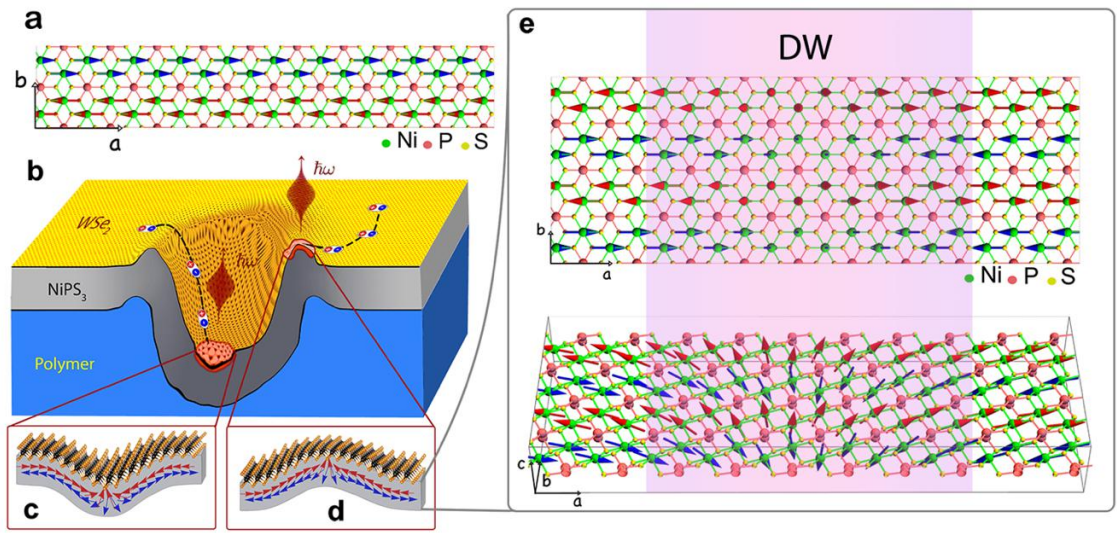

**Figure 3. Origin of chiral quantum light emission. a** Zig-zag AFM order overlayed with atomic NiPS$_3$ structure where Ni electronic spins are aligned parallel (blue arrows) and antiparallel (red arrows) to the crystallographic *a*-axis. **b** Schematic of an indentation region cross-section where local strain creates exciton capture centers in WSe$_2$. These centers can act as QEs of circularly polarized photons. **c, d.** DWs within strained regions of the AFM layer adjacent to the QEs result in the local ferromagnetic proximity effects breaking the time reversal symmetry within WSe$_2$. **e.** Atomic structure of top NiPS$_3$ layer illustrating top and side views of a DW (pink shaded region) giving rise to proximity effects.

Recently however, studies have revealed that defects in AFM order, *e.g.* domain walls (DWs), can create localized magnetizations.[45-47] A simple theoretical model (Supporting Information S1) shows that an AFM DW can carry uncompensated spin. Pinning of an AFM DWs in the confinement potential created by localized strain field of a nano-indentation would lead to the rotation of Néel vector and a localized out-of-plane magnetization. As our estimate of the local dipolar magnetic field due to DW magnetization (Figure 3e) is only of the order of 4-50 millitesla, which is orders of magnitude lower than ~1-10 T field required to induce the observed DCP (Supporting Information S2 and S3), we hypothesize that proximity effects, e.g., exchange interaction between AFM and localized exciton of TMD layer facilitate the time reversal symmetry breaking necessary for the emission of chiral single photons. Because the strain field of our nano-indentation is irregular, multiple local magnetizations with opposite direction may arise within a single indented region. As illustrated in Figure 3b-d, multiple independent QEs form within a single, locally strained indentation could couple to neighboring localized magnetizations of varying direction and strength. As a result, QEs with differing orientation and DCP emerge as reported in Figure 1 & 2. It is important to note that nano-indentation here plays a critical role in creating a QE within the monolayer WSe$_2$ which is colocalized to an AFM defect in the NiPS$_3$.

Experimental evidence that the localized strain distribution associated with nano-indentations leads to the localized ferromagnetic proximity effect is provided by our analysis of linear polarization of PL emission. Together with the emission form the WSe$_2$ monolayer, our PL spectra show a peak at 1.47 eV characterized by a narrow spectral linewidth and a strong linear polarization (DLP = $(I_s - I_p)/(I_s + I_p)$) of 0.76 (Extended Data Fig.2). Notably the DLP of this peak from an indented region (Extended Data Fig. 6) reduced drastically to 0.24. A closer look at $\sigma^{+/-}$ PL spectra (Extended Data Fig. 6a inset) reveal the peak is $\sigma^+$ polarized with DCP of 0.4. As recent studies have associated this linearly polarized, anisotropic excitonic emission to the emergence of AFM order in NiPS$_3$ with the Néel vector perpendicular to the direction of linear polarization,[11-13] a drastic reduction in DLP accompanied by the emergence of some DCP in exciton emission is evidence that indented regions could contain multiple domains with reoriented Néel vectors: a prerequisite for the emergence of localized magnetization and ferromagnetic proximity effects.

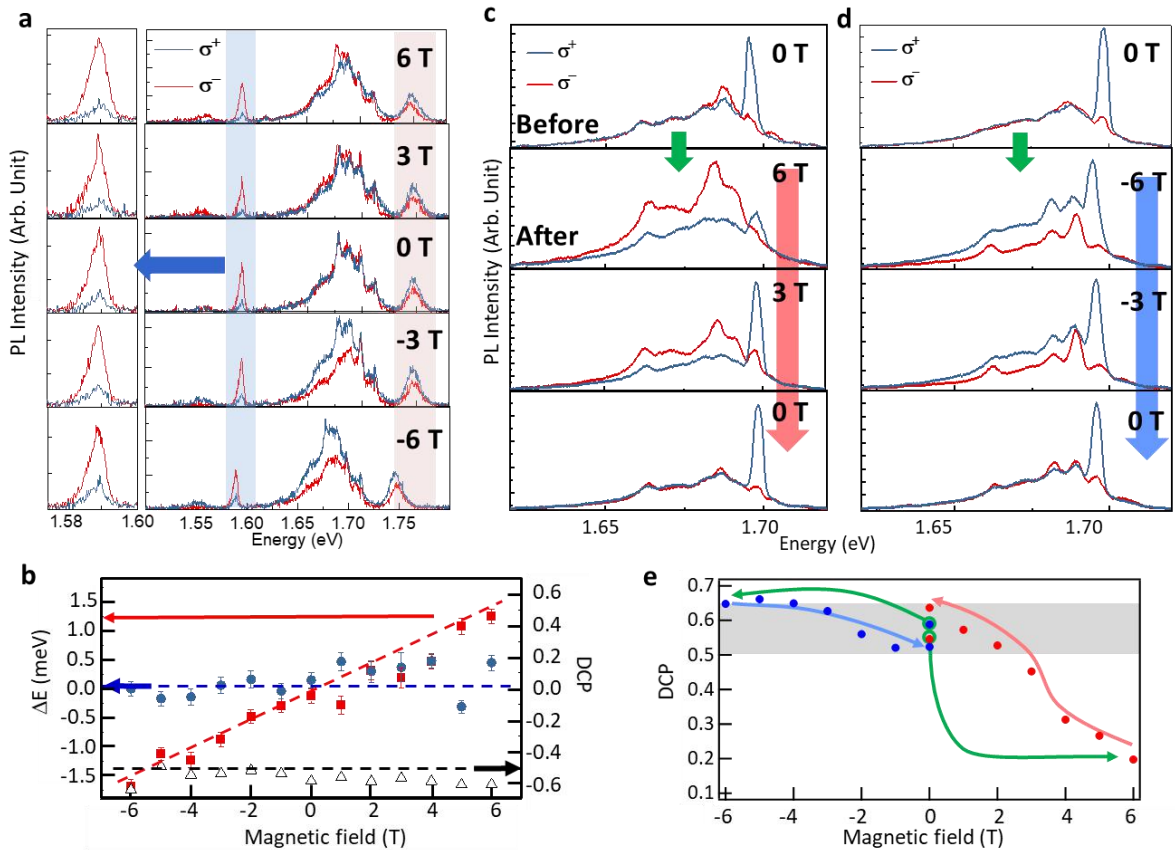

**Figure 4. Magneto-PL studies of chiral quantum emitters. a.** Low temperature, polarization-resolved PL spectra acquired as the function of external B field applied perpendicular to the sample plane (Faraday geometry). **Left panel**: Expanded view of a localized PL peak (1.59 eV) displaying strong DCP of 0.6. **b.** Energy splitting (left axis) between $\sigma^+$ and $\sigma^-$ PL peaks of localized exciton at 1.59 eV (blue data points) and 2D exciton at 1.75 eV (red data points). The DCP (right axis) of the 1.59 eV vs B field is plotted as the black diamonds. The blue, red, and black dash lines provide guides to the eye. **c. Top panel:** Low temperature, polarization resolved PL spectra of an indented region with a PL peak at ~1.70 eV exhibiting strong DCP of ~ 0.7 before the temperature and field dependent measurement. **2nd panels**: Low temperature, polarization resolved, PL spectra of the same location acquired after heating the sample to 180 K, raising the field to 6T and cooling the sample back down to 4K while maintaining B at 6T. **Bottom 2 panels**: Low temperature, polarization resolved, PL spectra acquired as the field is reduced back to zero T. **d.** Experiment in **c** was repeated by applying an external field of -6T at 180K. **e.** DCP of 1.69 eV PL peak plotted against external B field. Green circles and arrows represent the DCP of the PL peak originally observed in the topmost panel of **(c & d)** and the process involving heating the sample to 180K and cooling back down to 4K while maintaining the B field at 6 (-6T). Red (blue) arrows and data points represent the process and DCP observed upon reduction of B from 6T (-6T) to 0T at 4K immediately after cooling down. Since the DCP values are calculated from the peak intensities of $\sigma^{+/-}$ polarized spectra, they are contributed not only by the QE emission but also by the broad PL background that increase with decrease of field from 6T to -6T.

Studies of $NiPS_3$ revealed that external fields > 15T are required to reorient the Néel vector to the direction of magnetic field at cryogenic temperatures below the Néel transition temperature (150K).[13] Strain-induced localized magnetization associated with AFM defects are therefore expected to be stable against external B fields of similar strength. We are able to manifest clear experimental evidence of this stability using magneto PL spectroscopy. Here, polarization resolved PL spectra were acquired from an indentation under external B fields ranging from -6 T to +6T in the Faraday geometry **(Figure 4a,b** and Extended Data Fig.7 ). These PL spectra reveal that a strongly circularly polarized PL peak at ~1.59 eV with a DCP of 0.6 exhibits no observable Zeeman splitting between $\sigma^+$ and $\sigma^-$ emission peaks and no significant change in DCP (Figure 4b). B field invariance in both PL emission energy and the DCP provides evidence that robust AFM defects within $NiPS_3$ are responsible for the chiral quantum light emission. In marked contrast with the localized QE emission, the 2D $WSe_2$ exciton peak at 1.75 eV displays normal Zeeman splitting (~1.68 meV at ± 6 T) similar to uncoupled monolayers suggesting that proximity interactions average out to zero for delocalized 2D excitons, and that chiral PL emission demands co-localization of exciton and AFM defects.

The AFM Néel vector and magnetization of the AFM DWs, on the other hand, can be reoriented at a lower field (e.g. 6 T) if the field is applied at temperatures above the Néel transition temperature. We conducted a temperature dependent magneto PL experiment to test this hypothesis. First, a QE emitting strongly circularly polarized PL at 4 K and B= 0T was identified (1.7eV PL peak in Figure 4c). We then biased the spins of $NiPS_3$ to point out of the plane by raising the temperature above the Néel transition temperature to 180 K and increasing the B field to 6T along the sample normal. The sample was then cooled back to 4K while holding the B-field at 6T to maintain the spin orientation. PL spectra acquired after cooling to 4K at 6T (Figure 4c, 2$^{nd}$ panel from the top) showed a dramatic decrease of DCP of the QE to ~0.2 from original value of ~0.7, indicating that the local magnetization responsible for the QE's chiral PL emission is reduced by forcing the spins of $NiPS_3$ to align in the antiparallel direction. When the external B field is decreased, the DCP was found to increase until it was fully restored to its original value at 0T (Figure 4c,e, and Extended Data Fig. 7 & 8) suggesting that the local strain distribution restores the original local magnetization in absence of an external field. When a B field in opposite direction (i.e. -6T) is applied at 180 K, no significant change in DCP of the QE is observed since the spins of $NiPS_3$ are forced to align parallel to the local magnetization making it more stable as demonstrated in Figure 4d & e. This experiment provides strong evidence that our chiral QEs form due to proximity interactions with localized out-of-plane magnetizations associated with AFM defects in $NiPS_3$.

In summary, our observations reveal that local strain engineering can be utilized not only to create QEs, but also to localize ferromagnetic proximity effects required for creation of chiral single photon emitters in WSe$_2$/NiPS$_3$ heterostructures. These discoveries establish TMD/TMPX$_3$ AFM insulators as an exciting material platform for the further exploration of novel emergent phenomena and the realization of solid-state quantum transduction and sensing technologies.

**Methods.**

**Sample Preparation**

A polymethyl methacrylate solution was spin coated on Si substrate with a 285 nm SiO$_2$ layer to achieve a 300 nm thick uniform membrane. NiPS$_3$ flakes were exfoliated from the bulk crystals (2D Semiconductor) using Scotch tape and deposited onto the PMMA layer. Monolayers of WSe$_2$ (HQ Graphene) were mechanically exfoliated onto a silicone Gel-Film® stamp. The layer thickness was identified by optical microcopy contrast and room temperature PL. WSe$_2$ thin layers were then transferred onto the selected NiPS$_3$ flakes using a 2D material transfer system from HQ Graphene. NiPS$_3$ and WSe$_2$ exfoliations as well as layer stacking were performed in an argon glove-box to minimize degradation. Once the heterostructure was fabricated, nano-indentations were deterministically created by using a Bruker Dimension Icon atomic force microscope using blunt AFM probes with a nominal tip radius of 100 nm and a spring constant of 80 N/m. Indentations were fabricated by performing approach curves with a maximum force setpoint of between 50-60 μN. This combination of probe radius and force set-point was found to consistently produce chiral QEs within the WSe$_2$/NiPS$_3$ heterostructures. Topography imaging of the surface following nanoindentation was performed using tapping mode imaging with a second, topography probe (tip radius ~10 nm).

**Low Temperature Polarization Resolved PL Spectroscopy and Hanbury-Brown and Twiss Experiments.**

The experimental set up for optical measurements is schematically illustrated in Extended Data Fig. 1. The sample was held at 4K in a continuous flow cryostat (Oxford, MicrostatHiRes). The heterostructures were excited with a 514 nm CW laser (Coherent Obis) and PL signal is collected through a 50X, 0.55 NA Olympus microscope objective. A quarter-wave plate (QWP) is placed immediately after the objective to circularly polarized the laser excitation and convert the $\sigma^+$ and $\sigma^-$ polarized PL signal to S and P linear polarization. A Wollaston prism was then used to split the signal into channels which are then projected onto liquid nitrogen cooled CCD detector through a 500 mm spectrometer (Acton SpectraPro). This arrangement allows to collect spectral images of $\sigma^+$ and $\sigma^-$ PL signal on two different portion of the CCD camera simultaneously as shown in Extended Data Fig. 1. The $\sigma^+$ and $\sigma^-$ PL spectra and the extracted DCP are therefore free of artifacts that could result from random intensity fluctuation and spectral wandering. For linear polarization analysis, the QWP was removed and two half-wave plates (HWP1 & HWP2) were inserted in the excitation and the collection channels before the Wollaston prism. PL spectra are acquired as the function of half-wave plate rotation angles to determine the degree and orientation of PL linear polarization. HWP2 was replaced with a QWP to acquire $\sigma^+$ and $\sigma^-$ polarized PL spectra under linearly polarized excitation (Extended Data Fig. 5). For the time-tagged time correlated PL measurements, we pumped the sample with a tunable visible picosecond laser (TOPTICA Photonics, 40 MHz repetition rate, 3.5 ps pulse width) filtered by a 10 nm bandpass filter centered at 514 nm. The emission was then filtered by using a tunable band-pass filter (SLI) and coupled into an HBT spectrometer composed of a 50/50 beam-splitter and two avalanche photodiodes (MPD PDM Series) with timing accuracy of 35 ps. The macro and micro times of each photon detection event are recorded with a Hydraharp 400 TCSPC module and PL intensity time traces, PL lifetime and 2$^{nd}$ order photon correlation were extracted from the recorded photon stream.

**Temperature-dependent magneto-photoluminescence experiment**

To perform the magneto-photoluminescence experiments, the sample was mounted in the variable-temperature insert of a 7T magneto-optical cryostat (Oxford Instruments Spectramag). An in-situ aspheric lens (NA=0.68), mounted on piezoelectric nanopositioners (Attocube), focused the incident light. The sample was excited using a continuous-wave 632.8 nm HeNe laser. The polarization of the laser was controlled using a quarter-wave plate and linear polarizer. Emission from the sample in the Faraday geometry was collected by the same aspheric lens, and its polarization was analyzed using a quarter wave plate and linear polarizers. The emission was dispersed in a 500 mm spectrometer (Acton) and detected by a liquid nitrogen cooled CCD detector. For the temperature dependent magneto-PL measurements, we first set the sample temperature to 180 K (above the Néel temperature of the NiPS$_3$), then ramped the magnetic field to either +6T or -6 T. The sample was then cooled back to 4 K to repeat the field ramping experiments.

## Acknowledgements

This work was performed at the Center for Integrated Nanotechnologies, an Office of Science User Facility operated for the U.S. Department of Energy (DOE) Office of Science (OS). Los Alamos National Laboratory (LANL), an affirmative action equal opportunity employer, is managed by Triad National Security, LLC for the U.S. Department of Energy's NNSA, under contract 89233218CNA000001. Laboratory Directed Research and Development (LDRD) program 20200104DR provided primary support for the works of XL, HZ, AP, NS, SAC and HH. ACJ and VC acknowledge support by DOE BES, QIS Infrastructure Development Program, Deterministic Placement and Integration of Quantum Defects. XL, HH and SC also acknowledge partial support by Quantum Science Center, a National QIS Research Center supported by DOE, OS. JC and HZ also acknowledge a partial support from LANL Director's Postdoctoral Fellow Award. MTP acknowledges support from LDRD awards 20210782ER and 20210640ECR. The National High Magnetic Field Laboratory is supported by National Science Foundation (NSF) DMR-1644779, the State of Florida, and the U.S. Department of Energy (DOE).

## Author Contributions

HH conceived and led the experiment. XL fabricate $WSe_2/NiPS_3$ heterostructures, and discover chiral QEs through low temperature polarization resolved PL spectroscopy and HBT experiments. ACJ guided XL in creating nano-indentations in $WSe_2/NiPS_3$ heterostructure and scanning probe microscopy studies. SAC contributed the key idea of temperature dependent magneto-PL experiment and XL and JC conducted the experiment. HZ, MTP, and VC assisted XL in a variety of optical spectroscopy experiments. NS and AP provided the theoretical model. HH and XL wrote the manuscript with the assistance of all the authors.

**Corresponding Author**

Han Htoon, htoon@lanl.gov

**Extended Data**

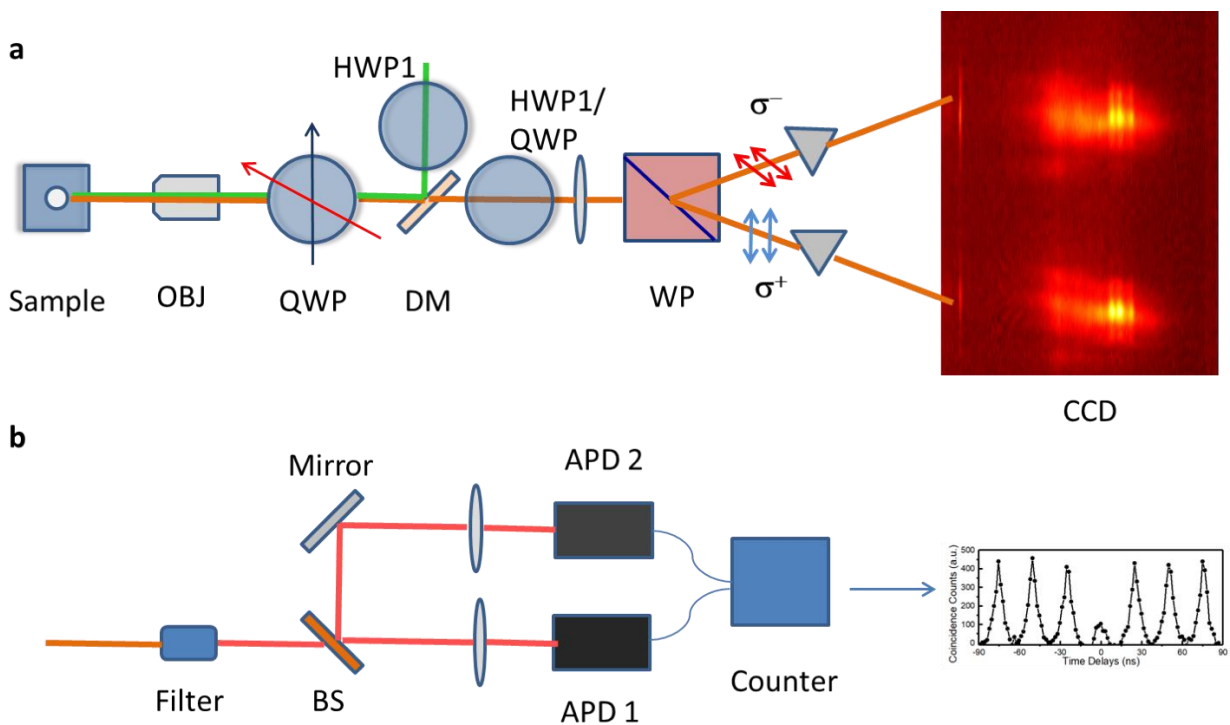

**Extended Data Fig. 1.a. Experimental set up for low temperature polarization resolved PL spectroscopy.** OBJ: Objective, DM: Dichroic mirror, QWP : quarter-wave plate, WP: Wollaston prism. For the linearly polarization measurements, QWP is removed and two half-wave plates (HWP1 and HWP2) are inserted in excitation and collection channels. **b. Schematic of HBT experiment.** BS: 50/50 beam splitter, APD1, APD2: Avalanche photo diodes, Counter: Hydra400 TCSPC module.

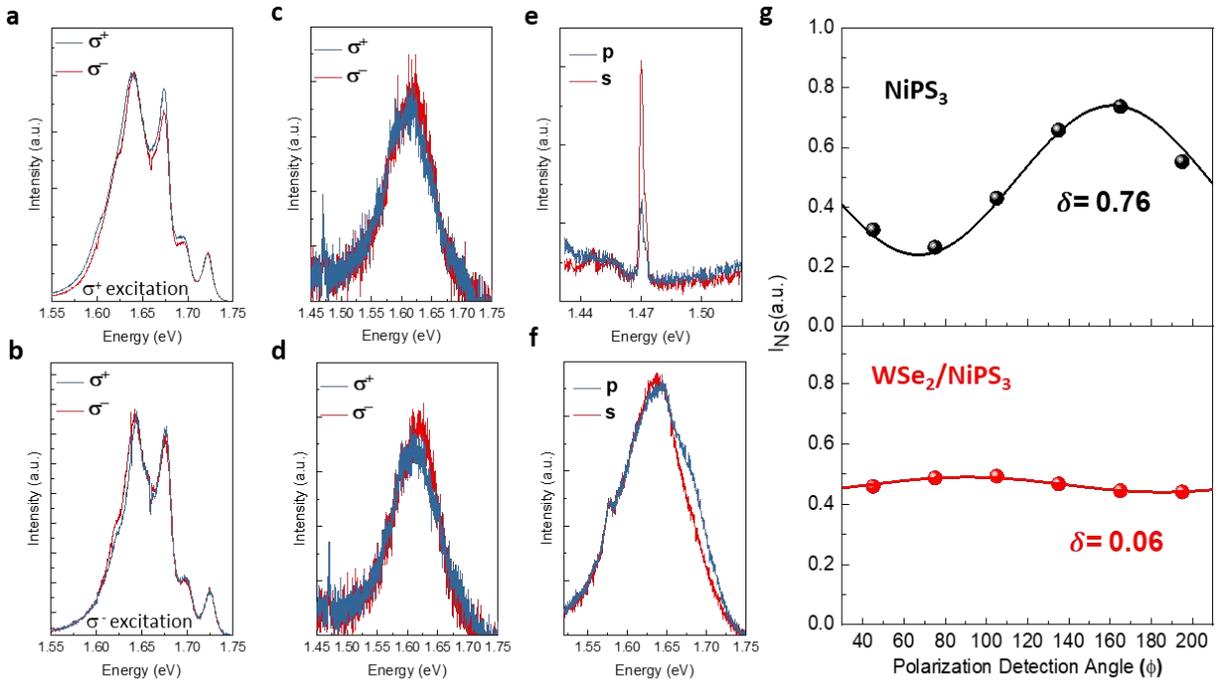

**Extended Data Fig. 2. Control experiment. a-d.** Low temperature, σ$^{+/-}$ polarized PL spectra of WSe$_2$ monolayer **(a, b)** and un-indented WSe$_2$/NiPS$_3$ heterostructure **(c, d)** excited by σ$^+$ **(a, c)** and σ$^-$ **(b, d)** 514 nm laser. The data show no detectable DCP in emission of both monolayer and un-indented WSe$_2$/NiPS$_3$ hetrostructure indicating that spin polarized excitons injected at 514 nm were depolarized completely upon relaxation to the band-edge. **e, f.** Low temperature, linearly (S, P) polarized PL spectra (left panel) and plot of DLP as the function of polarization detection angle (right panel) for the sharp emission peak at 1.47 eV **(e)** and WSe$_2$ exciton emission **(f)**. **(g) Upper Panel.** Strong DLP (0.76) observed for 1.47 eV sharp emission peak is consistent with prior studies that attributed the emission to highly anisotropic excitons reflecting the emergence of AFM order in NiPS$_3$. **Lower Panel.** PL of unstrained WSe$_2$/NiPS$_3$ heterostructure show weak degree of linear polarization consistent with behavior of WSe$_2$ monolayer.

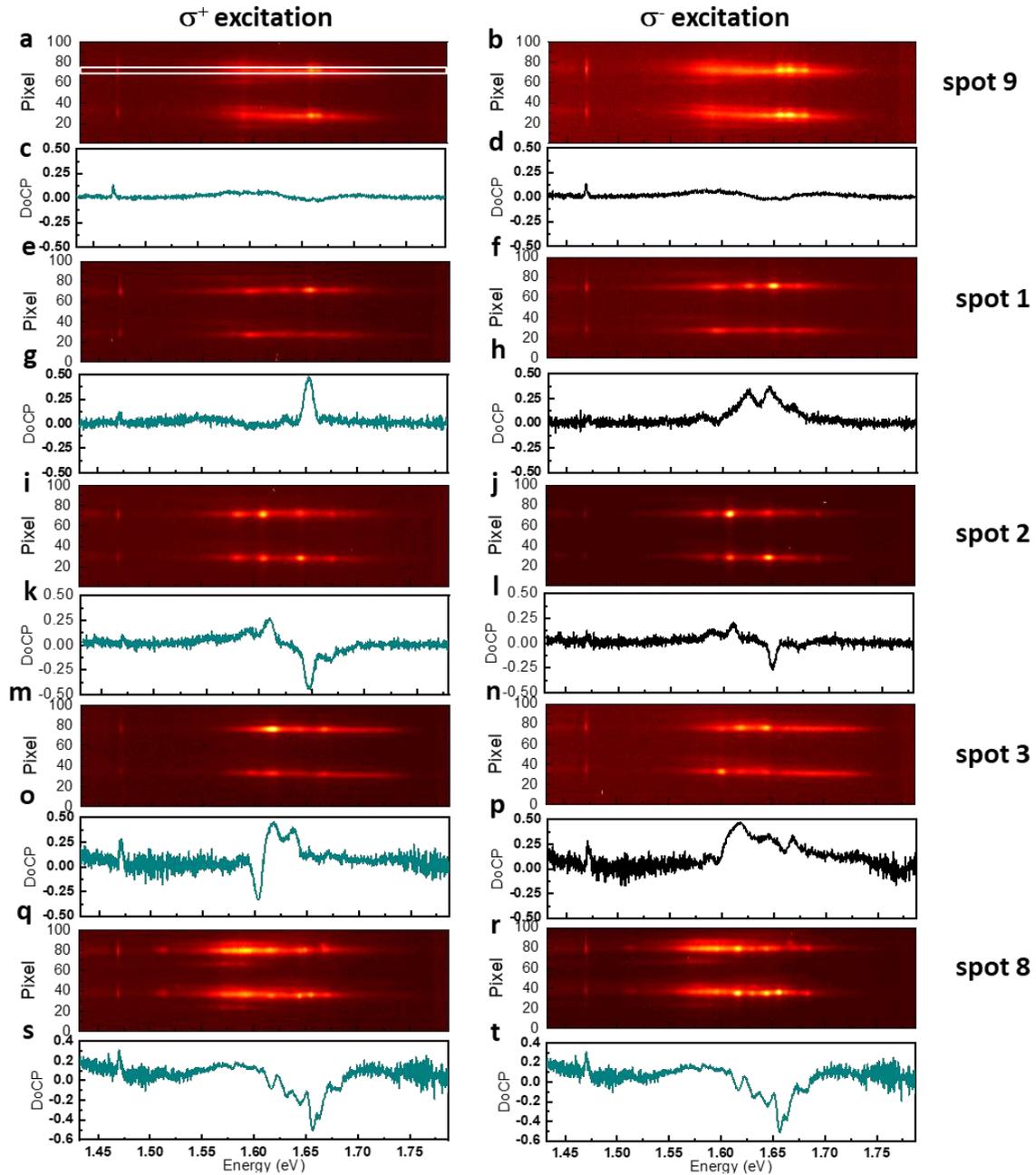

**Extended Data Fig. 3. PL spectral images and DCP spectra of Figure 1. a, b.** PL spectral images of σ⁺ and σ⁻ detection channels (Extended Data Fig. 1) recorded on upper and lower portion of the CCD detector for spot 9 under σ⁺ (a) and σ⁻ (b) laser excitation. **c, d.** DCP spectra for center 5 pixel (indicated by white rectangle) of spectral images (a) and (b). **e-h, i-l, m-p, q-t.** Spectral images and DCP spectra for spot 1, 2, 3, and 8, respectively. Simultaneous detection of σ⁺ and σ⁻ PL channels at two different region of CCD allow determination of DCP free of artifacts due to random temporal fluctuation of PL spectra.

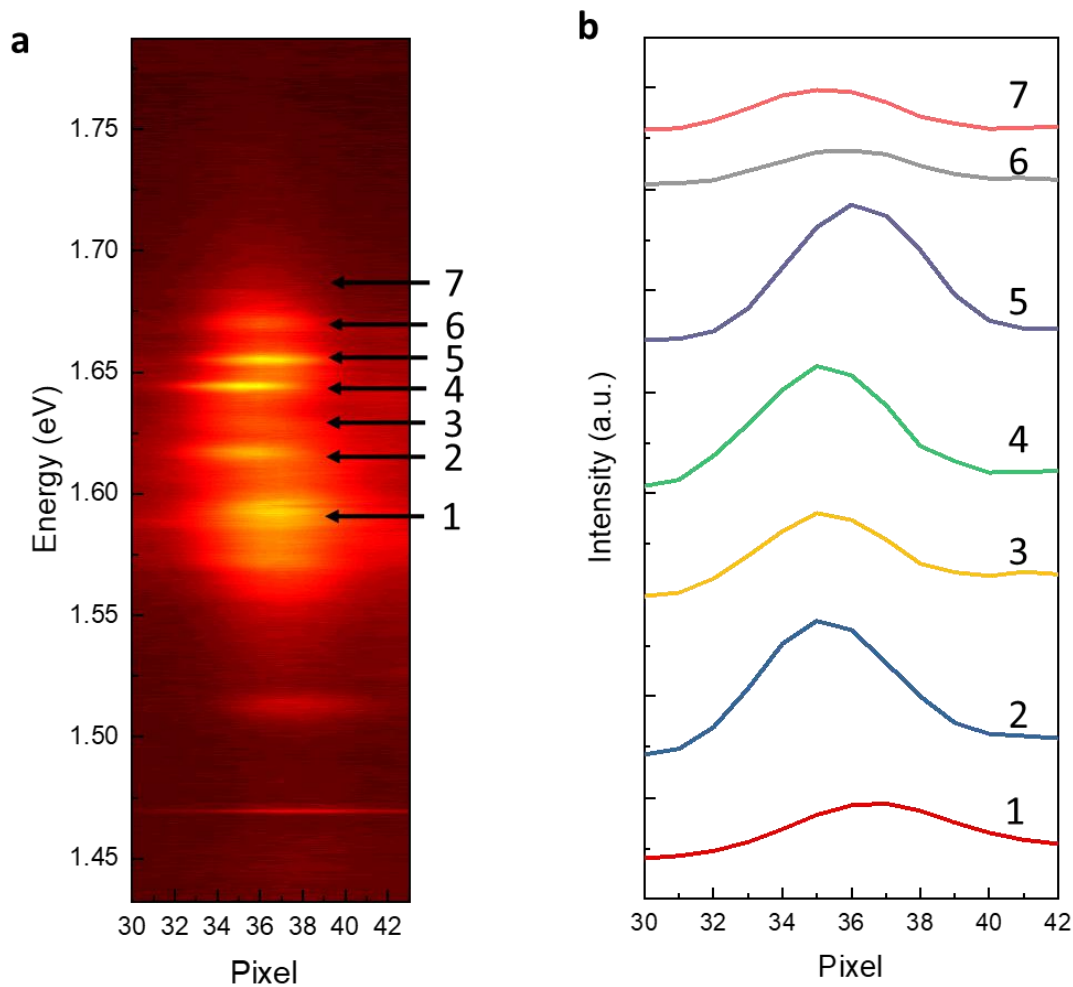

**Extended Data Fig. 4. Magnified spectral image and analysis of the spatial distribution of PL intensity. a.** Magnified spectra image of spot 3. **b.** Intensity distribution of the localized emission peaks marked 1-7 in a. By fitting the PL intensity distribution with Gaussian function, we can determine the centroid of the emission peak with ± 0.1 pixel precision. Based on the magnification factor of microscope objective (50X) and pixel dimensions of the CCD (20 μm), we can estimate that most of the sharp PL emission peaks originate from independent QE separated from one another by 50 to 100 nm.

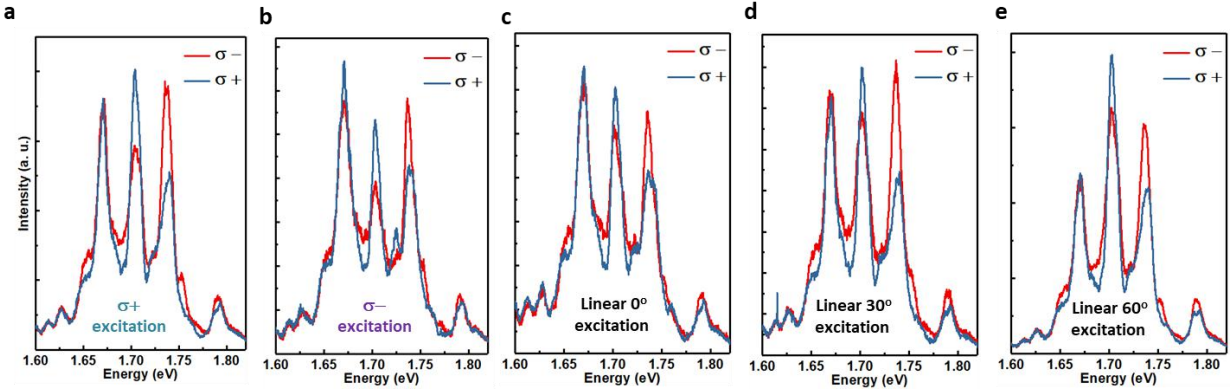

**Extended Data Fig 5. Invariance of circularly polarized PL emission under laser excitation of different polarization. a-e.** σ⁺ and σ⁻ polarized PL spectra of an indentation acquired under 514 nm laser excitation polarized at σ⁺ (a), at σ⁻(b), linearly at 0° (c), linearly at 30° (d) linearly at 60° (e). The PL peaks at 1.75 and 1.71 eV remain σ⁻ and σ⁺ polarized under all different laser excitation polarization.

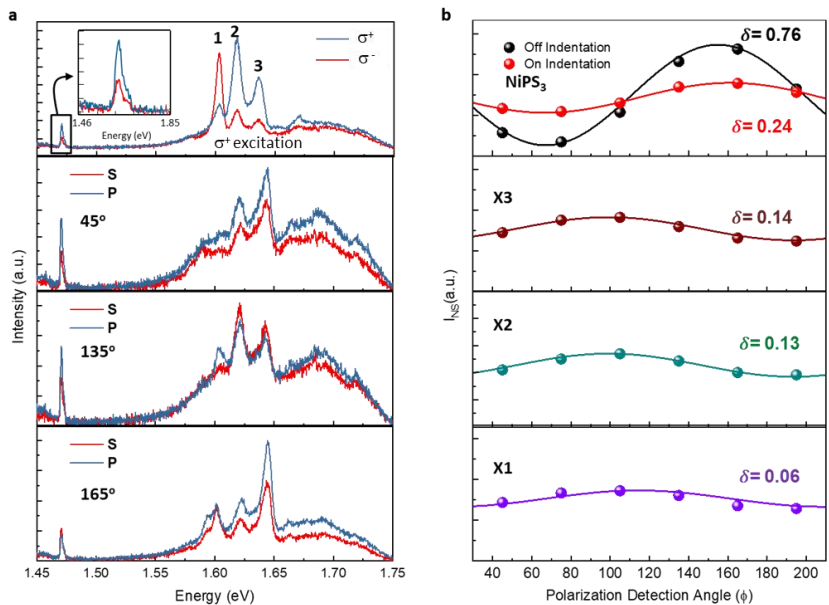

**Extended Data Fig. 6. Analyzing linear polarization of strain engineered WSe$_2$/NiPS$_3$ heterostructure. a, top panel.** σ⁺/σ⁻ polarized PL spectra of the spot 3 of Figure 1c acquired under σ⁺ excitation. **Inset.** Zoom-in view of the NiPS$_3$ anisotropic exciton peak displaying peak DCP of 0.4. **lower panels.** S and P linearly polarized PL spectra acquired at different polarization detection angles indicated in the panels. All three spectra are excited by σ⁺ polarized laser in the same way as the top panel. **b.** DLP of NiPS$_3$ anisotropic exciton (red data points and trace of **top panel**) and PL peak 1-3 (**lower panels**) as the function of polarization detection angle. Black data points and trace plot DLP of NiPS$_3$ anisotropic excitons in unstrained NiPS$_3$.

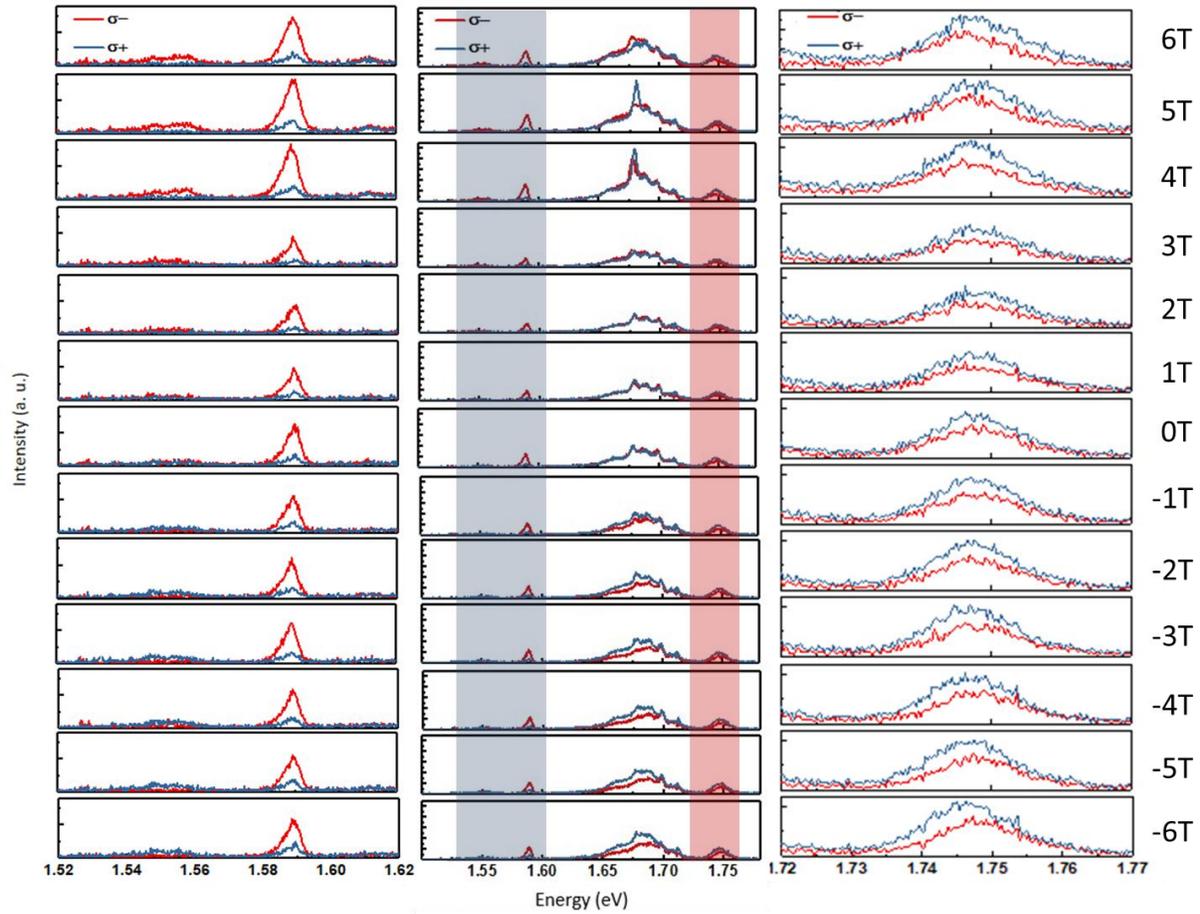

**Extended Data Fig. 7. Complete data set for Figure 3a. Center panel.** Low temperature σ⁺ and σ⁻ polarized PL spectra acquired under -6.0 T to +6.0 T external magnetic field applied perpendicular to the sample plane. **Left and right panels.** Zoom-in view of the gray and red spectra ranges of the center panels. Because magneto-PL spectra were acquired under 633 nm σ⁺ laser excitation, a weak DCP is observed at zero B field. A weak magnetic field dependent change in DCP likely resulting from field-induced polarization of localized excitons is also observed in 1.65 to 1.70 eV spectral range.

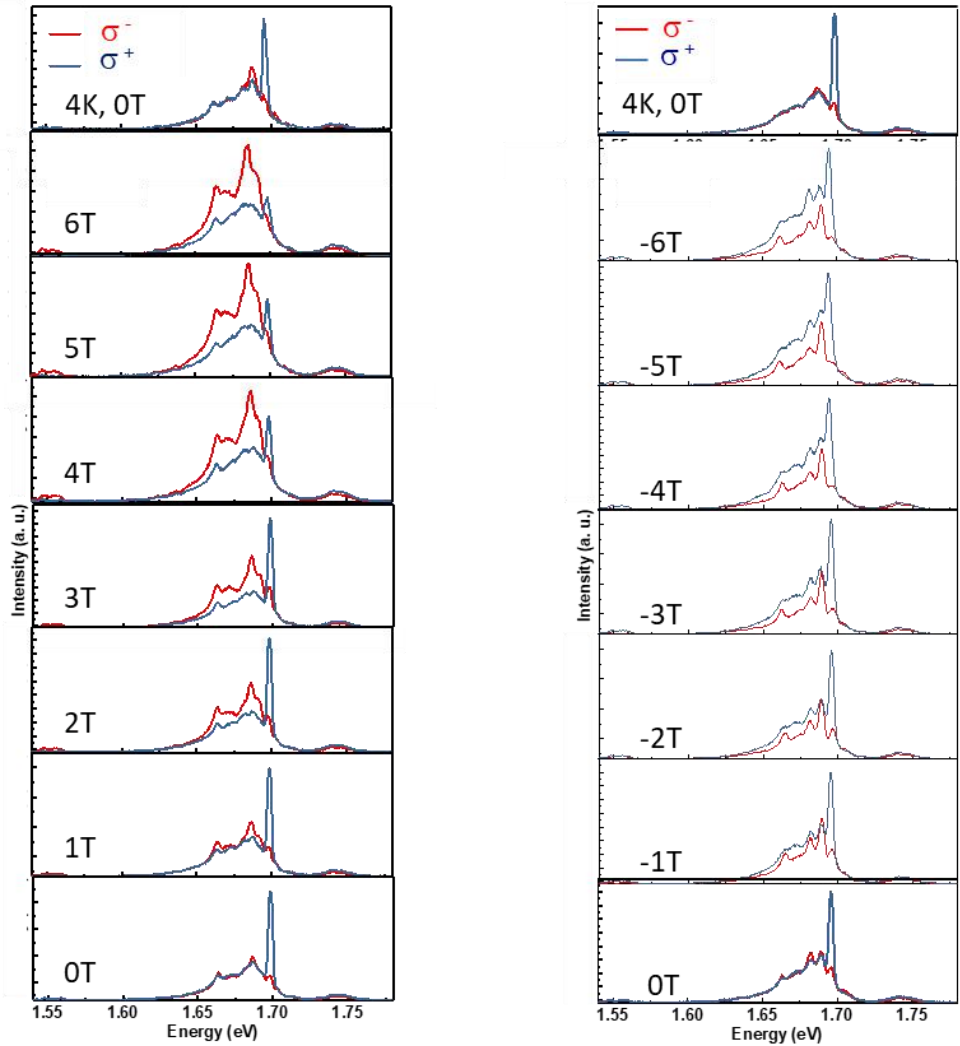

**Extended Data Fig. 8. Complete data set for figure 4c and 4d. a, b.** Complete B field dependent σ[+/−] polarized PL spectra shown in bottom 3 panels of Figure 4c and Figure 4d.

# Supporting Information

# Proximity Induced Chiral Quantum Light Generation in Strain-Engineered WSe$_2$/NiPS$_3$ Heterostructures

Xiangzhi Li,[1] Andrew C. Jones,[1] Junho Choi,[2] Huan Zhao,[1] Vigneshwaran Chandrasekaran,[1] Michael T. Pettes,[1] Andrei Piryatinski,[3] Nikolai Sinitsyn,[3] Scott Crooker,[2] Han Htoon[1*]

[1] Center for Integrated Nanotechnologies, Materials Physics and Applications Division, Los Alamos National Laboratory, Los Alamos, NM 87545

[2] National High Magnetic Field Laboratory, Materials Physics and Applications Division, Los Alamos National Laboratory, Los Alamos, NM 87545

[3] Theoretical Division, Los Alamos National Laboratory, Los Alamos, NM 87545

## S1 Breaking time-reversal invariance near defects in antiferromagnets

Antiferromagnetic ordering breaks the time-reversal invariance but this is usually hard to detect optically because oppositely oriented electronic spins generally compensate the magnetoelectric effects at distances larger than a lattice constant. For example, the magnetization averaged over a distance corresponding to the wavelength of light is practically zero. Let us show here that the situation can be different locally near defects, such as near mechanically induced local strains, that in turn may lead to local changes of a magnetic crystal anisotropy or pin spin-topological defects. Imagine that a local strain pins an anti-ferromagnetic domain wall. Such a wall generally carries an uncompensated spin. This fact is easy to verify using an easy axis anisotropy 1D Hamiltonian of classical spins:

$$H = \sum_{k=-L}^{L} \{J\vec{S}_{2k} \cdot \vec{S}_{2k+1} - D(S_{2k}^z)^2 - D(S_{2k+1}^z)^2\}, \quad J, D > 0 \qquad (1)$$

The positive value of $J$ forces nearby spins to point in opposite directions, and $D$ is the crystal anisotropy that makes the spins point along the z-axis. Here the index $k$ runs over all two-spin unit cells of the lattice with a large $L$. Let $\phi_s$ be the angle between the $s$-th spin and the z-axis. One ground state of (1) then corresponds to the sequence of phases

$$0, \pi, 0, \pi, \cdots, 0, \pi, 0, \pi,$$

where $\phi_s = 0$ corresponds to the spins with odd index and $\phi_s = \pi$ is for spins with even indices. The presence of the antiferromagnetic domain wall, which flips the direction of the Neel vector, is observed at a large distance as a flip of the phases of spins with the same index parity, e.g.,

$$0, \pi, 0, \pi, \cdots, \pi, 0, \pi, 0. \tag{2}$$

If we continue the latter sequence from both left and right to the center of the domain wall, one spin has to remain uncompensated, which means that the antiferromagnetic domain wall carries a finite magnetization. To explore its direction, consider the case of strong antiferromagnetic coupling: $J \gg D$. Then, the angle difference between the two spin vectors minus $\pi$ is small. The net z-polarization of two nearby spins with indices $2k$ and $2k + 1$ is then

$$S^z_{2k} + S^z_{2k+1} = S(\cos\phi_{2k} + \cos\phi_{2k+1}) = 2S\cos\left(\frac{\phi_{2k}+\phi_{2k+1}}{2}\right)\sin\left(\frac{\phi_{2k}-\phi_{2k+1}}{2}\right), \tag{3}$$

where $S$ is the spin size. For $J \gg D$, we use the continuum approximation

$$\phi_{2k+1} \approx \phi_{2k} + \pi + \frac{1}{2}\frac{d\phi_{2k}}{dz}dz,$$

where we assumed that the distance along z is measured so that the unit cell size (with two spins) is the unit of length, i.e., here $dz = 1$. For $J \gg D$, the derivative of phase is small, so (3) simplifies:

$$S^z_{2k} + S^z_{2k+1} \approx -\frac{S\sin\phi_{2k}}{2}\frac{d\phi_{2k}}{dz}dz.$$

Already here, we find that any gradient of the Neel vector, which corresponds to a nonzero $\frac{d\phi_{2k}}{dz}$, leads to a local spin polarization. Hence, in the vicinity of strains, such a polarization may not average to zero. Summing over all pairs of spins throughout the domain wall we find the net domain wall magnetization to be

$$M = -\int_{-L}^{L}\frac{S\sin\phi_{2k}}{2}\frac{d\phi_{2k}}{dz}dz = -\frac{S}{2}[\cos\phi_{2L} - \cos\phi_{-2L}], \tag{4}$$

where $\phi_{\pm 2L}$ are the phases far from the domain wall. For the sequence (2), the phases change from $\phi_{-2L} = \pi$ to $\phi_{2L} = 0$. Hence, the domain wall magnetization in this example points along the z-axis and has a magnitude

$$M = S, \tag{5}$$

that is, it equals the magnetization of one spin.

It is clear from this example that pinning of an antiferromagnetic domain wall by a local mechanical strain leads to emergence of a local magnetization and hence local breaking of the time-reversal symmetry. Although in this 1D example the net magnetization of the domain wall is microscopic, in a 3D sample, the domain wall is already a 2D surface that can produce a considerable local magneto-optical effect. Local strains can simply pin such domain walls by their strong local anisotropy field. However, the precise microscopic physics leading to the magnetooptical effects may be different. For example, a local strain can change the local direction of anisotropy, and thus lead to rotation of the Neel vector to an arbitrary angle. Thus, if the anisotropy axis changes from *z* to *x*, the Neel vector rotates between these two directions. It is easy to verify that this creates a local uncompensated spin polarization both along z and x directions. Therefore, a precise magnetization direction is determined by microscopic physics that distorts the Neel vector near the mechanical strain, and may not correspond to just the domain wall pinning.

All such microscopic effects in a large sample average to zero. For example, the domain walls in a macroscopic sample may carry opposite magnetization signs with equal chances. However, a specific local strain should be able to stabilize only one defect at a specific special point. Its magnetization then induces a ferromagnetic proximity effect in the nearby region of a TMD material, which should be observable as a circular-polarization dependence of the light absorption near this defect.

## S2 Magnetic field induced by domain wall in a layer of NiPS$_3$ antiferromagnet

In this section, we estimate magnetic field produced by a domain wall in NiPS$_3$ in the direction perpendicular to the surface of proximal TMD material and demonstrate that the magnetic dipole field at the distance range of 0.3 – 1.0 nm is already too weak to cause observable circular polarization. This suggests that proximality effects, e.g., exchange interaction between the strained antiferromagnetic substrate and strained TMD material enhances the local magnetic field facilitating circularly polarized photon emission as observed in experiment.

According to Ref.[S1], the Hamiltonian (1) describes formation of 1D domain wall in antiferromagnets. An important parameter determining the extent of such a domain wall over $L_{eff}$ lattice sites is

$$L_{eff} = \sqrt{J/D} \,. \tag{6}$$

If, $L_{eff} \sim 1$, the continuum approximation introduced in the previous section can be used. For NiPS$_3$, we adopt J = 5 meV and D = 0.3 meV, [S2] resulting in $L_{eff} = 4$ which gives us a crude justification for using the continuum approximation. Let us define Neel vector as

$$n_k = \frac{1}{2S}(S_{2k-1} - S_{2k}), \tag{7}$$

where $S = |S_{2k}|^2 = |S_{2k=1}|^2$ and dimensional magnetization density as

$$m_k = 4\mu_B(S_{2k-1} + S_{2k}), \tag{8}$$

with $\mu_B$ denoting Bohr magneton.

For an antiferromagnetic domain wall components of the Neel vector and magnetization density can be expressed as [S1]

$$n_k^x = -\tanh(k/L_{eff}), \quad n_k^y = 0, \quad n_k^z = \frac{1}{\cosh(k/L_{\text{eff}})}, \tag{9}$$

$$m_k^x = \frac{2\mu_B\sqrt{s(s+1)}}{L_{eff}} \frac{1}{\cosh^2(k/L_{\text{eff}})}, \quad m_k^y = 0, \quad m_k^z = \frac{\sqrt{s(s+1)}}{L_{eff}} \frac{\tanh(k/L_{eff})}{\cosh(k/L_{\text{eff}})}, \tag{10}$$

respectively.

The magnetic field at position $r$ induced by an array of magnetic dipoles occupying sites $r_k$ is

$$B(r) = \frac{\mu_o}{4\pi} \sum_k \left( \frac{3(r-r_k)((r-r_k)\cdot m_k)}{|r-r_k|^5} - \frac{m_k}{|r-r_k|^3} \right), \tag{11}$$

where $\mu_o$ denotes vacuum permeability and components of magnetization vector $m_k$ are given in (10).

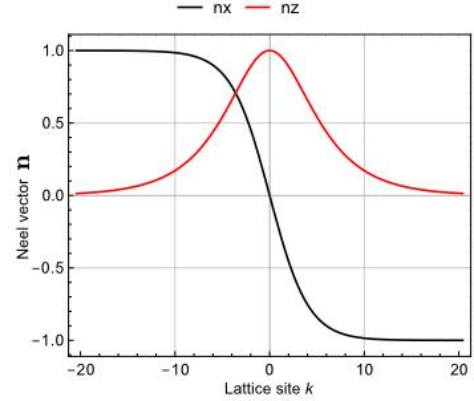

**Fig. S1** Calculated Neel vector components with, $n_x$ being within the plane of two-dimensional sheet of $Ni_2PS_3$ and $n_z$ perpendicular to the sheet.

Using the model above, we further perform calculations of domain wall within a 2D layer of $NiPS_3$ as illustrated in Fig. 3e. According to Fig. 3a, the antiferromagnetic structure in $NiPS_3$ is due to zig-zag configuration of Ni electronic spins (S=1) oriented parallel/antiparallel to the crystallographic vector $a$. [S2] To make an upper estimate for the magnetic field, we simplified the model by replacing the zig-zag configurations of interacting S=1 spins with a square lattice $(a, b)$ ($|a| = 0.58$ nm and $|b| = 1.00$ nm) of spins with $S = 2$. We further identified axis x and y to be oriented along the lattice $a$ and $b$ vectors, respectively. Accordingly, direction z is perpendicular to the $NiPS_3$ sheet.

Using (9) we calculated projections of Neel vector on x and z- axes as shown in Fig. S1. The plot indicates the spin flip within the domain wall. The HWHM of $n_z$ reflects effective domain wall delocalization length of $L_{eff} = 4$ unit cells as estimated above. The magnetic field in z-direction for various values of distance $z$ from the surface of $NiPS_3$ is calculated using (10) and (11) and shown in Fig. S2. The plots show that within the range of 0.35 nm < z < 1.00 nm, maximum of the magnetic field $B_z$ decreases from 0.08 T to 0.004 T. Such small values can be explained based on the observation made in the previous section that 1D domain wall creates a single uncompensated spin contributing to the magnetic field. Although we account for a series of uncompensated spins in the y-direction their contributions to the magnetic field drop as $r^{-3}$ because of the dipolar nature of the field. It has been reported in the literature and we provide our estimate below that the Zeeman splitting corresponding to the well pronounced circular polarization of the photons emitted by a localized quantum defect in $WSe_2$, requires magnetic field on the order of a few T. Our calculations show that the dipolar field is not enough to account for the dipolar magnetic field due to a single domain wall in NiPS$_3$ substrate to observe circularly polarized photon emission from proximal WSe$_2$. This discrepancy suggests that proximity interactions enhance the magnetic field to the values required for the obseration of circularly polarized photons.

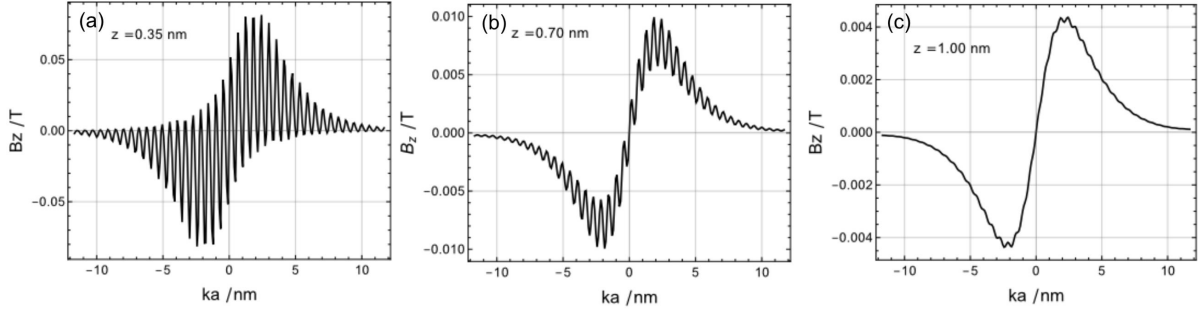

**Fig. S2** Calculated distribution of dipolar magnetic field induced by domain wall distance (a) z=0.35 nm, (b) z=0.70nm, and (c) z=1.00nm away from NiPS$_3$ surface.

**S3 Degree of circular polarization (DCP) vs applied external magnetic field**

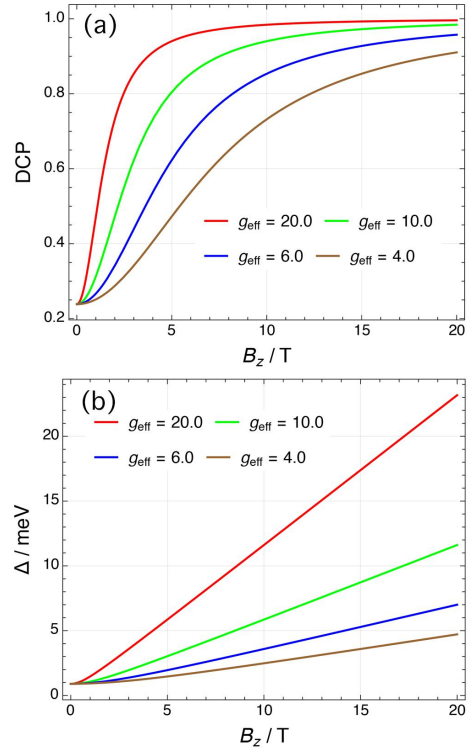

As defined in the main text

$$DCP = \frac{I_{\sigma_+} - I_{\sigma_-}}{I_{\sigma_+} + I_{\sigma_-}}. \quad (12)$$

Here, assuming Lorentzian emission line shapes of $\gamma_\pm$ widths, intensity of the right-handed and left-handed photons emitted within spectral interval $\omega_{min} \leq \omega \leq \omega_{max}$ are

$$I_{\sigma_+} \sim \int_{\omega_{min}}^{\omega_{max}} \frac{\gamma_+}{(\omega_+ - \omega)^2 + \gamma_+^2} d\omega, \quad (13)$$

$$I_{\sigma_-} \sim \int_{\omega_{min}}^{\omega_{max}} \frac{\gamma_-}{(\omega_+ + \Delta - \omega)^2 + \gamma_-^2} d\omega, \quad (14)$$

respectively. In (14), $\Delta = \omega_+ - \omega_-$ denotes Zeeman splitting between the right-handed and left-handed photon emission lines centered at $\omega_+$ and $\omega_-$, respectively. The Zeeman splitting can be estimated as [S3]

$$\Delta = \sqrt{\Delta_0^2 + (\mu_B g_{eff} B_z)^2}, \quad (15)$$

with $\Delta_0$ being zero-filed fine structure splitting, $\mu_B$ Bohr magneton, $g_{eff}$ effective g-factor, and $B_z$ magnetic field component perpendicular to the TMD plane.

**Fig. S3** Calculated (a) DCP and (b) related Zeeman splitting vs external magnetic field for different values of effective g-factor.

Using (12) – (15), we performed calculations of the DCP

as a function of external magnetic field $B_z$. The following parameters are adopted in our model: $\gamma_+ = \gamma_- = 1.0$ meV and $\Delta_0 = 0.9\ meV$. Effective g-factor depends on the electronic structure of emitting centers. Following Ref. [S4], we explore the following range of values for WSe$_2$ defects: the highest reported value due to ferromagnet proximity effect $g_{eff} = 20.0$, average value due to ferromagnet proximity effect overlapping with the upper boundary for uncoupled defects $g_{eff} = 10.0$, average value for uncoupled defects $g_{eff} = 6.0$, and the lower boundary for the uncoupled defects $g_{eff} = 4.0$. For DCP associated with the right-handed photon emission line, $\omega_{min} = \omega_+ - \gamma_+/2$ and $\omega_{max} = \omega_+ + \gamma_+/2$ in (13) and (14).

The results shown in Fig. S3a clearly show that for adopted $g_{eff} = 20 - 4.0$, an external magnetic field required to reach experimentally observed DCP ~ 0.5 – 0.7 (associated with $\Delta$ ~ 1.5 – 2.5 eV) should be on the order of few T. Taking into account that subtraction of the broad emission background observed in our experiment can boost DCP values close to one, Fig. S3 suggests that tens of T would be required to achieve Zeeman splitting of $\Delta \geq 5$ meV facilitating such DCP values. Compared to 0.004 T< B$_z$ < 0.08 T reported in the previous section for the dipolar field of a single domain wall in NiPS$_3$, our calculations rule out this scenario supporting the hypothesis of proximity effects causing large values of Zeeman splitting and DCP.